\documentstyle[12pt]{article}

\newcommand{\beq}{\begin{equation}}
\newcommand{\bea}{\begin{eqnarray}}
\newcommand{\D}{{\cal D}}
\newcommand{\dsl}
  {\kern.06em\hbox{\raise.15ex\hbox{$/$}\kern-.56em\hbox{$\partial$}}}
\newcommand{\id}
 {i\kern.06em\hbox{\raise.25ex\hbox{$/$}\kern-.60em$\partial$}}
\newcommand{\as}{/\kern-.52em s}
\newcommand{\bs}{/\kern-.52em t}
\newcommand{\eeq}{\end{equation}}
\newcommand{\eea}{\end{eqnarray}}
\newcommand{\ZZ}{{\rm \kern 0.275em Z \kern -0.92em Z}\;}
\newcommand{\Z}{{\cal Z}}
\begin{document}
\title     {Bosonization of Vector and Axial-Vector
                Currents in $3+1$ Dimensions.}
%
\author{%
C.D. Fosco\thanks{Investigador CONICET}\\
{\normalsize\it
  Centro At\'omico Bariloche}\\
{\normalsize\it
  8400 Bariloche, Argentina}\\
\rule{0cm}{1.cm}
F.A. Schaposnik\thanks{Investigador CICBA}\\
{\normalsize\it
  Departamento de F\'{\i}sica, Universidad Nacional de La Plata}\\
{\normalsize\it
  C.C.~67, (1900) La Plata, Argentina}%
}
\date{\today}
\maketitle
\begin{abstract}
The bosonization of a massless fermionic field coupled to both vector
and axial-vector external sources is developed, following a path-integral approach.
The resulting bosonized theory contains two antisymmetric tensor fields whose
actions consist of  non-local Kalb-Ramond-like terms plus interactions.
Exact bosonization rules that take the axial anomaly for the axial current into account
are derived, and an approximated bosonized action is constructed.
\end{abstract}
\newpage
Any physical system is suitable of many different mathematical
descriptions, i.e., there is a lot of freedom in the choice of
variables used to define its configurations. Different choices of
variables are equivalent in the sense that they describe the same
system. An extreme manifestation of this appears in some
two-dimensional models, which can be described in terms of either
fermionic or bosonic variables. The equivalence between these
two formulations  is made explicit by the so called  bosonization rules,
that map the fermionic to the bosonic variables.

There has been some progress in the program of extending, at least
in an approximated way, the bosonization procedure to theories
in more than two dimensions \cite{BQ}-\cite{LNS}.
In this letter we are concerned with the
problem of finding the bosonization rules for a
simple system in four space-time dimensions, that of a massless
fermionic field in the presence of vector and
axial sources. This will allow us to find the bosonization rules
for the vector and axial fermionic currents, this last one being
consistent
with the axial anomaly~\cite{anom1} in the presence of an external
vector gauge field.

Our starting point is the generating functional for a massless  Dirac
field in  $3+1$ (Euclidean) dimensions, coupled to Abelian vector ($s_\mu$)
and axial-vector ($t_\mu$) external sources
$$\Z ( s_\mu , t_\mu ) \;=\; \int \D {\bar \psi} \D \psi \,
\exp \left[ - S( {\bar \psi} , \psi ; s_\mu , t_\mu ) \right]$$
\beq
S( {\bar \psi} , \psi ; s_\mu , t_\mu ) \;=\; -i  \int d^4 x \; {\bar \psi}
\, (\id  - \as  -  \gamma_5 \bs ) \, \psi   \;
\label{strt}
\eeq
where we have adopted for the $\gamma$-matrices the following conventions:
\beq
\gamma_\mu^\dagger \;=\; \gamma_\mu \;\;,\;\;
\gamma_5^\dagger \;=\; \gamma_5 \;\;,\;\;
\{ \gamma_\mu \,,\,\gamma_\nu \} \,=\, 2 \, \delta_{\mu\nu} \;.
\eeq

The addition of the source $s_\mu$ is motivated by the reason that, 
in four dimensions, the vector and axial currents are {\em independent \/}
fermionic bilinears. Thus not all the information provided by 
$\Z (s_\mu, t_\mu)$ can be obtained from, say, $\Z (s_\mu , 0)$. 
In two dimensions, because of the smaller number of generators for
the Dirac algebra, these two current are related, and the bosonization
rule for one of the currents also implies the proper rule for the other.

By a natural extension of the procedure followed to obtain the bosonized
version of the generating functional with just a vector source
(see for example \cite{FS}-\cite{LNS} for a detailed description of
the approach) , we perform
in (\ref{strt}) the change of variables
\beq
\psi (x) \,=\, e^{i \theta (x) - i \gamma_5 \alpha (x) } \psi' (x) \;,\;
{\bar \psi} (x)\,=\,{\bar \psi}' (x) e^{- i \theta (x) - i \gamma_5 \alpha(x)}
\;.
\label{trafo}
\eeq
In terms of the new variables, the generating functional (\ref{strt}) reads
\beq
\Z ( s_\mu , t_\mu ) \;=\; \int \D {\bar \psi} \D \psi \,
J (\alpha ; s_\mu , t_\mu) \,
\exp \left[ - S( {\bar \psi} , \psi \;;\; s_\mu + \partial_\mu
\theta , t_\mu + \partial_\mu \alpha) \right]
\label{scnd}
\eeq
where the primed fermionic fields have been renamed as unprimed for
the sake of simplicity, and $J$ is the anomalous Jacobian corresponding to
this fermionic change of variables, a well-known consequence of the chiral
anomaly~\cite{jaco}. This Jacobian is evaluated by using the standard
Fujikawa's
recipe. To decide whether the consistent or covariant regularization is to be used in the
calculation of this determinant, it is important to recall (see
for example ref.\cite{Fuj2}) that the consistent regularization, besides  assuring the
validity of Wess-Zumino consistency conditions, automatically enforces the
conservation of the vector current. The covariant regularization, in turn,
guarantees a gauge-covariant form for the anomaly, at the expense of
generating an anomaly for the vector current as well.
Regarding the case at hand, it is desirable, although not mandatory, to
have a regularization that assures the conservation of the vector
current, since one has in mind applications of the bosonization recipe to
situations where interactions involve the vector current, which should
then be non-anomalous, while the axial current can have an anomalous
divergence without spoiling the consistency of the theory.  This
sets the natural choice of regularization in this case to be the consistent
one, which we adopt. This justifies {\em a posteriori\/}
the fact that we have written $J$ in (\ref{scnd}) as a function of
$\alpha$ only, rather than depending also on $\theta$.
To obtain the Jacobian for this finite transformation, we shall use the
techniques described in ref.\cite{fjac}. The consistent regularization of the
anomalous Jacobian may be based on the hermitean operator
\beq
{\cal D} \;=\; i \not \! \partial \,-\, \not \! s \,+\, i \gamma_5
\not \! t
\eeq
obtained by performing the analytic continuation $t_\mu \to -i t_\mu$ in
the operator appearing in the kinetic term of the
fermionic action. One then
introduces a parameter $r \in [0,1]$, to define the $r$-dependent
transformation
\bea
\psi (x) &=& e^{- i r \theta (x) - r \alpha (x) \gamma_5  }\; {\psi'}_r (x)
\nonumber\\
{\bar \psi} (x) &=& {{\bar \psi}'}_r (x) e^{ i r \theta (x) - r \gamma_5
\alpha(x)}
\label{rtrafo}
\eea
which, of course, induces a change in the operator ${\cal D}$
$$
{\cal D}_r \;=\; e^{ i r \theta (x) - r \gamma_5 \alpha(x)}
{\cal D} e^{- i r \theta (x) - r \alpha (x) \gamma_5  }
$$
\beq
=\; i \not \! \partial \,-\, ( \not \! s - r \not \! \partial \theta )
\,+\, i \gamma_5 ( \not \! t - r \not \! \partial \alpha ) \;.
\eeq
The Jacobian $J$ for the finite transformation (\ref{trafo})
corresponds to $r = 1$, and may be obtained by multiplying
the Jacobians corresponding to infinitesimal steps $dr$, each one
calculated with the appropriated regularization operator ${\cal D}_r$
\beq
J \;=\; \exp \left\{ 2 \int_0^1 dr \, \lim_{t \to 0+} \, {\rm Tr}
[ \alpha \gamma_5 e^{- t {\cal D}^2_r} ] \right\} \;.
\label{jact}
\eeq
The factor $2$ in the exponent comes from multiplying the
two Jacobians,
associated to  the changes in the measures $\D \psi$ and $\D {\bar \psi}$.
Although each one of these Jacobians does depend on $\theta$, like
terms do cancel in the product. This is a consequence of the
consistent regularization, which uses the same regularization operator
for both $\D \psi$ and $\D {\bar \psi}$.

There just remains to insert into (\ref{jact}) the result for the
functional trace, which can be read from the formula for the
anomaly for an infinitesimal transformation
\beq
\lim_{t \to 0+} \, {\rm Tr}
\left[ \alpha \gamma_5 e^{- t {\cal D}^2_r} \right]
\;=\;
- \frac{1}{8 \pi^2} \epsilon_{\mu\nu \rho\sigma}
(- \partial_\mu s_\nu \partial_\rho s_\sigma \,+\,
\frac{1}{3} \, \partial_\mu t_\nu \partial_\rho t_\sigma )
\eeq
which, in this Abelian case, becomes $r$-independent, rendering
the integration over $r$ in (\ref{jact}) trivial.
After undoing the analytic continuation on $t$,
the anomalous Jacobian for the transformation (\ref{trafo}) becomes
\beq
J (\alpha ; s_\mu , t_\mu)\;=\; \exp \left[ \frac{1}{4 \pi^2} \int
d^4 x \, \alpha (x) \, \epsilon_{\mu\nu\rho\sigma}
(\partial_\mu s_\nu \partial_\rho s_\sigma \,+\, \frac{1}{3}
\partial_\mu t_\nu \partial_\rho t_\sigma)\right] \;.
\label{jaco}
\eeq
The next step in the bosonization procedure follows from realizing that,
as (\ref{trafo}) is a change of variables, $\Z$ cannot depend on either
$\theta$ or $\alpha$. Thus $\theta$ and $\alpha$ can be integrated out
without other effect than the introduction of  an irrelevant constant
factor in $\Z$, which we ignore. Integration over $\theta$ and $\alpha$ is
equivalent to integration over two flat Abelian vector fields
$\theta_\mu$ and $\alpha_\mu$:

$$\theta_\mu \;=\; \partial_\mu \theta \;\;\; ,
\;\;\; \alpha_\mu \;=\; \partial_\mu \alpha$$
\beq
f_{\mu\nu}(\theta) \; \equiv \; \partial_\mu \theta_\nu - \partial_\nu \theta_\mu \,=\,0
                           \;\;,\;\;
f_{\mu\nu}(\alpha) \; \equiv \; \partial_\mu \alpha_\nu - \partial_\nu \alpha_\mu \,=\,0 \;.
\label{defs}
\eeq
The fermionic action obviously depends on $\alpha$ and $\theta$ only through
$\alpha_\mu$ and $\theta_\mu$ defined in (\ref{defs}).
By  an integration by parts, one sees that so does the Jacobian (\ref{jaco}).
Had the situation been different (i.e., through
the dependence of $J$ on $\alpha$
and not only on its gradient $\alpha_\mu$), it would not  be possible to pass
from a description in terms of $\alpha$ and $\theta$ to one in terms of
$\alpha_\mu$ and $\theta_\mu$.
To make this explicit, it is convenient to define
\beq
J(\alpha_\mu ; s_\mu , t_\mu)\;=\; \exp \left[ - \frac{1}{4 \pi^2} \int
d^4 x \, \alpha_\mu (x) \, \epsilon_{\mu\nu\rho\sigma}
(s_\nu \partial_\rho s_\sigma \,+\, \frac{1}{3}
t_\nu \partial_\rho t_\sigma)\right] \;.
\label{jaco1}
\eeq
Thus (\ref{scnd}) becomes
$$
\Z \;=\; \int \D \theta_\mu \, \D \alpha_\mu \,\D {\bar \psi}\,\D \psi \;
\delta [f_{\mu\nu}(\theta)] \,
\delta [f_{\mu\nu} (\alpha)]\;
J(\alpha_\mu ; s_\mu , t_\mu)
$$
\beq
\exp \left[ - S( {\bar \psi} , \psi ; s_\mu + \theta_\mu , t_\mu +
\alpha_\mu ) \right] \;.
\label{trd}
\eeq
Formally integrating out the fermionic fields and making the shift of variables
\beq
\theta_\mu \; \to \; \theta_\mu \,-\, s_\mu \;\;\;,\;\;\;
\alpha_\mu \; \to \; \alpha_\mu \,-\, t_\mu \; ,
\eeq
(\ref{trd}) leads to
$$
\Z (s_\mu,t_\mu)\,=\, \int \D \theta_\mu \, \D \alpha_\mu \,\delta[f_{\mu\nu}(\theta
-s)] \delta[f_{\mu\nu}(\alpha-t)]\, J(\alpha_\mu - t_\mu  ; s_\mu , t_\mu)
$$
\beq
\times \det ( \not \! \partial + i \not \! \theta + i \gamma_5 \not
\! \alpha) \;.
\label{frth}
\eeq
By analogy with the bosonization procedure followed in
\cite{FS}-\cite{LNS}, we exponentiate the
two functional delta functions in (\ref{frth}) using two antisymmetric tensor fields
$A_{\mu\nu}$ and $B_{\mu\nu}$ as Lagrange multipliers
$$ \Z (s_\mu, t_\mu) \;=\; \int \D A_{\mu\nu} \, \D B_{\mu\nu} \,
\D \theta_\mu \, \D \alpha_\mu \;\; J(\alpha_\mu - t_\mu  ; s_\mu , t_\mu)$$
$$ \exp\left({i \int d^4 x [\epsilon_{\mu\nu\rho\sigma}
A_{\mu\nu} (\partial_\rho \theta_\sigma - \partial_\rho s_\sigma) \;+\;
\epsilon_{\mu\nu\rho\sigma} B_{\mu\nu}  ( \partial_\rho \alpha_\sigma -
\partial_\rho t_\sigma ) ]}\right) \;$$
\beq
\times \;\; \det ( \not \! \partial + i \not \! \theta +
i \gamma_5 \not \! \alpha) \;.
\label{fvth}
\eeq
The bosonized form of $\Z$ can then be obtained by integrating out
$\theta_\mu$ and $\alpha_\mu$ in (\ref{fvth}). This produces
a generating functional with the tensor fields $A_{\mu\nu}$ and
$B_{\mu\nu}$ as dynamical variables. This step requires
the evaluation of the fermionic determinant, which in four dimensions
is necessarily non-exact.
Before embarking on such calculation, we derive the rules that
map the vector and axial-vector currents into functions of the
bosonic fields $A_{\mu\nu}$ and $B_{\mu\nu}$. This
correspondence requires no approximation and may well be called `exact'.
These rules follow from elementary functional differentiation
\beq
j_{\mu} = \langle {\bar \psi} \gamma_\mu \psi \rangle
= -i \frac{\delta}{\delta s_\mu} \log \Z |_{s_\mu = 0}
= - \epsilon_{\mu\nu\rho\sigma} \partial_\nu A_{\rho\sigma}
\label{ve}
\eeq
\beq
j_{5 \mu} = \langle {\bar \psi} \gamma_5 \gamma_\mu \psi \rangle
= -i \frac{\delta}{\delta t_\mu} \log \Z |_{t_\mu = 0}
= - \epsilon_{\mu\nu\rho\sigma} \partial_\nu B_{\rho\sigma}
- \frac{i}{4\pi^2} \, \epsilon_{\mu\nu\rho\sigma}
s_\nu \partial_\rho s_\sigma \;.
\label{ax}
\eeq
From the antisymmetry of the tensors $A_{\mu\nu}$ and $B_{\mu\nu}$,
we are entitled to derive the equations for the divergencies of the
currents:
\bea
\partial_\mu j_\mu &=& 0 \nonumber\\
\partial_\mu j^5_\mu &=& - \frac{i}{8 \pi^2} \,
{\tilde F}_{\mu\nu}(s) F_{\mu\nu}(s) \;.
\eea
with
${\tilde F}_{\mu\nu} =(1/2) \epsilon_{\mu\nu\alpha\beta}F_{\alpha\beta}$.
We then see that the bosonization rule (\ref{ax}) correctly reproduces
the axial anomaly.

As stated above, although the bosonization recipe (\ref{ve})-(\ref{ax})
for associating the fermionic currents with expressions written
in terms of bosonic fields is exact, the bosonic action governing
the boson field dynamics cannot be evaluated in an exact form in
$d>2$ dimensions. Different approximations for computing the
fermionic determinant would yield alternative effective
bosonic actions valid in different regimes.
We shall  describe here the evaluation of the fermionic determinant
in (\ref{fvth}) to second order in the fields $\theta_\mu$ and
$\alpha_\mu$. The use of this quadratic approximation can be
motivated by the same kind of arguments (see in particular the
`quasi-theorem') used in ref.~\cite{bos1}.

As usual, it is convenient to work in terms of
$W$, the generating functional of connected Green's functions of
the fermionic current
\beq
\det (\not \! \partial \, + \, i \not \! \theta \,+\, i \gamma_5
\not \! \alpha) \;=\; \exp \left[ W (\theta_\mu , \alpha_\mu) \right] \; .
\label{defw}
\eeq
The {\em unregularized \/} form of the quadratic part of $W$ in  (\ref{defw})
becomes
\beq
W (\theta_\mu , \alpha_\mu) \;\simeq\; \frac{1}{2} \,
{\rm Tr} \, \left[
\not \! \partial^{-1} (\not \! \theta + \gamma_5 \not \! \alpha )
\not \! \partial^{-1} (\not \! \theta + \gamma_5 \not \! \alpha )
\right] \;.
\eeq
$W (\theta_\mu , \alpha_\mu)$ will consist of three parts
\beq
W (\theta_\mu , \alpha_\mu) \;=\;
W_{\theta\theta} (\theta_\mu , \theta_\mu) \;+\;
W_{\alpha\alpha} (\alpha_\mu , \alpha_\mu) \;+\;
W_{\theta\alpha} (\theta_\mu , \alpha_\mu)
\label{split}
\eeq
corresponding to the three terms in the quadratic part of
$W$. Before extracting the three parts of $W$, a regularization should be
introduced, since, at it will become clear next, it does have a non-trivial effect.
Our choice of a regularization is dictated by the requisite of compatibility
with the regularization of the anomalous Jacobian, which we decided to
be consistent. A form of assuring consistency is to use Pauli-Villars
regularization in a way that treats the vector and axial-vector vertices
symmetrically. This amounts to defining the regulated $W$ by
\beq
W_{reg} ( \theta_\mu , \alpha_\mu ) \;=\; \frac{1}{2} \,
\sum_{s=0}^2 C_s \,  {\rm Tr} \,
\left[ (\not \! \partial + M_s)^{-1} \, (\not \! \theta + \gamma_5 \not \! \alpha)
(\not \! \partial + M_s)^{-1} \, (\not \! \theta + \gamma_5 \not \! \alpha)\right]
\eeq
where
$$
C_0 \,=\, 1 \;\;\; C_1 \,=\, 1 \;\;\;   C_2 \,=\, -2 \;
$$
\beq
M_0^2  \,=\,\mu^2 \;\;\;M_1^2 \,=\, 2 \Lambda^2 - \mu^2 \;\;\; M_2^2 \,=\,\Lambda^2
\eeq
where $\Lambda$ is the cutoff and we have introduced an IR regulator $\mu$ that
gives the gauge field a small mass inside the loop. The regulated version of
(\ref{split}) becomes
\beq
W_{reg} (\theta_\mu , \alpha_\mu) \;=\;
W_{reg}^{\theta \theta} (\theta_\mu , \theta_\mu) \;+\;
W_{reg}^{\alpha\alpha} (\alpha_\mu , \alpha_\mu) \;+\;
W_{reg}^{\theta\alpha} (\theta_\mu , \alpha_\mu)
\eeq
with
\beq
W_{reg}^{\theta \theta} (\theta_\mu , \theta_\mu) \;=\;
\frac{1}{2} \sum_{s=0}^2 \, C_s {\rm Tr} \,
\left[
(\not \! \partial + M_s)^{-1} \not \! \theta \,
(\not \! \partial + M_s)^{-1} \, \not \! \theta
\right]
\label{wtt}
\eeq
\beq
W_{reg}^{\alpha\alpha} (\alpha_\mu , \alpha_\mu) \;=\;
\frac{1}{2} \sum_{s=0}^2 \, C_s {\rm Tr} \,
\left[
(\not \! \partial + M_s)^{-1} \not \! \alpha \,
(\not \! \partial - M_s)^{-1} \not \! \alpha
\right]
\label{waa}
\eeq
for the first two terms. Knowledge of the mixed term involving
$\theta_\mu$ and $\alpha_\mu$ is not necessary, since because of
parity violation and Lorentz invariance its form is restricted to be
\beq
W_{reg}^{\theta \alpha} (\theta_\mu , \alpha_\mu)
\;=\; \int d^4 x \int d^4 y \; \epsilon_{\mu\nu\rho\sigma} \,
\partial_\mu \theta_\nu (x)
H (x - y)
\partial_\rho \alpha_\sigma (y)
\eeq
which is a total derivative, and can thus be safely ignored (at least in this
{\rm Abelian} case we are dealing with). We then consider the terms (\ref{wtt})
and (\ref{waa}).
The first one is evidently equivalent to the quadratic part of the effective
action for a Pauli-Villars regulated fermionic field in the presence of an
external vector field $\theta_\mu$. Setting the renormalization conditions
at zero momentum, we have for this renormalized two-point function
\beq
W_{reg}^{\theta \theta} (\theta_\mu , \theta_\mu)  \,=\,
\frac{1}{2} \int d^4 x d^4 y \, \theta_\mu (x) \delta^\perp_{\mu\nu} F (x - y)
\theta_\nu (y)
\eeq
where $\delta^\perp_{\mu\nu}$ is the transverse $\delta$
\beq
\delta^\perp_{\mu\nu} \;=\; \delta_{\mu\nu} \,-\, \partial_\mu \partial^{-2}
\partial_\nu
\eeq
and
$$
F(x-y) \;=\; \int \frac{d^4 k}{(2 \pi)^4} \, e^{i k \cdot (x-y)}
{\tilde F} (k)
$$
\beq
{\tilde F} (k) \,=\, \frac{k^2}{2 \pi^2} \,\int_0^1 dx \,x (1-x)
\, \log \left[ 1 + x (1-x) \frac{k^2}{\mu^2} \right] \;.
\label{deff}
\eeq
As it will become evident at the end, to impose a renormalization condition
on $W^{\theta \theta}$ is equivalent to set {\em the same} renormalization
condition on the (one-particle irreducible) current-current correlation
function. Of course, the current is the vector one for $\theta_\mu$ and
the axial-vector one for $\alpha$.

Regarding the term (\ref{waa}), we may rearrange it in such a way that
it shows a part which is functionally identical to (\ref{wtt}), but with
$\alpha_\mu$ as argument, plus a remainder $\Delta W_{reg}^{\alpha\alpha}$
\beq
W_{reg}^{\alpha\alpha} (\alpha_\mu, \alpha_\mu) \;=\;
W_{reg}^{\theta\theta} (\alpha_\mu, \alpha_\mu) \;+\; \Delta W_{reg}^{\alpha\alpha}
\label{wdw}
\eeq
where
\beq
\Delta W_{reg}^{\alpha\alpha} \;=\;
\sum_{s=0}^{s=2} \, C_s M_s^2 \, {\rm Tr} \,
\left[ (-\partial^2  + M_s^2)^{-1}
\not \! \alpha (- \partial^2 + M_s^2)^{-1} \not \! \alpha \right] \;.
\eeq

Of course, the  renormalized term $W^{\theta\theta} (\alpha_\mu, \alpha_\mu)$
will be given by
\beq
W^{\theta \theta} (\alpha_\mu , \alpha_\mu)  \,=\,
\frac{1}{2} \int d^4 x d^4 y \, \alpha_\mu (x) \delta^\perp_{\mu\nu} F (x - y)
\alpha_\nu (y)
\eeq
with $F$ as in (\ref{deff}).

A straightforward evaluation shows that $\Delta W_{reg}^{\alpha\alpha}$
is given by
\beq
\Delta W_{reg}^{\alpha\alpha} \;=\; - \frac{\Lambda^2}{(2 \pi)^2}
\int d^4 x \, \alpha_\mu (x) \alpha_\mu (x) \;.
\eeq
The meaning of this quadratic divergence is that the consistent
regularization violates axial gauge invariance, as expected.
The renormalization of this divergence requires the introduction of a
mass counterterm, and also the choice of renormalization conditions
for $W^{\alpha \alpha}$ which should say which is the value of
the renormalized mass for $\alpha$. Thus the renormalized
$\Delta W^{\alpha\alpha}$ will correspond to a finite mass term:
$$
\Delta W^{\alpha\alpha} \;=\; - \, \frac{m^2}{2}
\int d^4 x \, \alpha_\mu (x) \alpha_\mu (x)
$$
\beq
= \;- \frac{m^2}{2}
\int d^4 x \left[ \alpha_\mu (x) \delta^\perp_{\mu\nu} \alpha_\mu (x)
\,+\, \alpha_\mu (x) \delta^\parallel_{\mu\nu} \alpha_\mu (x) \right]
\eeq
where $\delta^\parallel_{\mu\nu} =  \partial_\mu \partial^{-2}
\partial_\nu$, and $m$ is the renormalized mass. Of course we can
set the value of the renormalized mass to zero to this order. If
$\alpha_\mu$ is absolutely non-dynamical, it is possible to set
the renormalized mass equal to zero to all orders, since the only
correction to the mass term is produced by the term we are considering.
If the field were dynamical, the anomalous behaviour of the axial symmetry
would have spoiled the masslessness of $\alpha$ to higher orders.
From the previous results it follows that we can write the renormalized
functional $W$ as follows:
$$
W (\theta_\mu , \alpha_\mu) \;=\;-\frac{1}{2} \, \int d^4 x d^4 y \,
\left[ \theta_\mu (x) \delta^\perp_{\mu\nu} F(x-y) \theta_\nu (y)
\right.
$$
\beq
\left. +\,\alpha_\mu (x) \delta^\perp_{\mu\nu} G(x-y) \alpha_\nu (y)
\,+\, m^2 \, \alpha_\mu (x) \delta^\parallel_{\mu\nu} \delta (x-y)
\alpha_\nu (y) \right]
\label{wren}
\eeq
where
\beq
G(x-y) \;=\; F(x-y) \,+\, m^2 \delta (x-y) \;.
\eeq

Inserting (\ref{wren}) into (\ref{fvth}), we see that the functional
integral is Gaussian with respect to $\theta_\mu$ and $\alpha_\mu$:
$$ \Z (s_\mu,t_\mu) \,=\, \int \D A_{\mu\nu} \, \D B_{\mu\nu} \,
\D \theta_\mu \, \D \alpha_\mu $$
$$
\exp \left[ - \frac{1}{4 \pi^2} \int d^4 x \, (\alpha_\mu  - t_\mu) \,
\epsilon_{\mu\nu\rho\sigma} (s_\nu \partial_\rho s_\sigma \,+\, \frac{1}{3}
t_\nu \partial_\rho t_\sigma)\right]
$$
$$ 
\exp \left\{ 
i \int d^4 x [\epsilon_{\mu\nu\rho\sigma}
A_{\mu\nu} (\partial_\rho \theta_\sigma - \partial_\rho s_\sigma) \;+\;
\epsilon_{\mu\nu\rho\sigma} B_{\mu\nu}  ( \partial_\rho \alpha_\sigma -
\partial_\rho t_\sigma ) ]\right\}
$$
$$
\exp \left\{ -\frac{1}{2} \, \int d^4 x d^4 y \,
[ \theta_\mu (x) \delta^\perp_{\mu\nu} F(x-y) \theta_\nu (y)\right.
$$
\beq
\left. + \,\alpha_\mu (x) \delta^\perp_{\mu\nu} G(x-y) \alpha_\nu (y)
\,+\, m^2 \, \alpha_\mu (x) \delta^\parallel_{\mu\nu} \delta (x-y)
\alpha_\nu (y) ]
\right\}
\;.
\label{sxth}
\eeq

Performing the Gaussian integration over $\theta_\mu$ and
$\alpha_\mu$ in (\ref{sxth}) we get
\begin{eqnarray}
& & \Z (s_\mu,t_\mu) \,=\, \exp [{\cal C} (s_\mu,t_\mu)] \,
\int \D A_{\mu\nu} \, \D B_{\mu\nu} \times \nonumber\\
& & \exp \left\{-i \int d^4 x [ s_\mu \epsilon_{\mu\nu\rho\sigma}
\partial_\nu A_{\rho\sigma} + t_\mu (\epsilon_{\mu\nu\rho\sigma}
\partial_\nu B_{\rho\sigma}+ \frac{i}{4 \pi^2}\epsilon_{\mu\nu\rho\sigma}
s_\nu \partial_\rho s_\sigma)] \right\} \times \nonumber\\
& &
\exp \left\{ - \frac{1}{3} \, \int d^4 x d^4 y [ A_{\mu\nu\rho}(x)
F^{-1}(x-y) A_{\mu\nu\rho}(y) + \right.\nonumber \\
& & \left. B_{\mu\nu\rho}(x)
G^{-1}(x-y) B_{\mu\nu\rho}(y) ] \right\} \times
\exp \left\{ - \frac{i}{4 \pi^2} \int d^4 x d^4 y
\partial_\mu B_{\nu\rho}(x) \times \right. \nonumber\\
& & \left. G^{-1}(x-y) \delta_{\mu\nu\rho ,
\alpha\beta\gamma} (s_\alpha \partial_\beta s_\gamma + \frac{1}{3}
t_\alpha \partial_\beta t_\gamma) \right\}
\label{boson}
\end{eqnarray}
where
\bea
A_{\mu\nu\rho} &=& \partial_\mu A_{\nu\rho} + \partial_\nu A_{\rho\mu} +
\partial_\rho A_{\mu\nu} \nonumber\\
B_{\mu\nu\rho} &=& \partial_\mu B_{\nu\rho} + \partial_\nu B_{\rho\mu} +
\partial_\rho B_{\mu\nu} \nonumber\\
\delta_{\mu\nu\rho ,\alpha\beta\gamma}&=&
\det \left( \begin{array}{ccc}
\delta_{\mu\alpha} & \delta_{\mu\beta} & \delta_{\mu\gamma}\\
\delta_{\nu\alpha} & \delta_{\nu\beta} & \delta_{\nu\gamma}\\
\delta_{\rho\alpha} & \delta_{\rho\beta} & \delta_{\rho\gamma}
\end{array}
\right)
\eea
and
$$
{\cal C} (s_\mu,t_\mu) \,=\, \frac{1}{2 (2 \pi)^4} \int d^4 x d^4 y \,\left\{
[ s_\mu (x) \partial_\nu s_\lambda (x)\,+\, \frac{1}{3} t_\mu (x)\partial_\nu
t_\lambda (x) ] \right.
$$
$$
\delta_{\mu\nu\rho ,\alpha\beta\gamma} G^{-1} (x-y)
[ s_\alpha (y) \partial_\beta s_\gamma (y)\,+\, \frac{1}{3} t_\alpha (y)\partial_\beta
t_\gamma (y) ]
$$
$$
+ \frac{1}{2 (2\pi)^4} \int d^4 x d^4 y \, {\cal G} (x)
\partial^{-2} G^{-1} (x-y) {\cal G} (y)
$$
\beq
\left. + \frac{1}{2 m^2 (2 \pi)^4} \int d^4 x d^4 y \,{\cal G}(x)
\partial^{-2} (x-y) {\cal G} (y)\right\}
\label{chu}
\eeq
where ${\cal G} = \epsilon_{\mu\nu\rho\lambda}( \partial_\mu  s_\nu
\partial_\rho
s_\lambda \,+\, \frac{1}{3} \partial_\mu t_\nu \partial_\rho t_\lambda)$.
\newpage
\section*{Summary and Conclusions}
We have applied the bosonization technique developed in 
ref's~[5-9] to the case of massless Dirac fermions in
four dimensions in the presence of both vector and axial-vector 
sources. This has allowed us to find the bosonization rules 
for both fermionic currents, eqs.(\ref{ve})-(\ref{ax}),
in terms of Kalb-Ramond bosonic
fields. While the bosonization rule for the vector current
can be written in a natural and compact form, reminiscent
of the well-known two-dimensional bosonization rule, 
\beq
\bar \psi \gamma_\mu \psi \to 
-\epsilon_{\mu \nu \rho \sigma} \partial_\nu A_{\rho \sigma} \, ,
\label{ult}
\eeq
the result for the axial current is more involved and includes
the vector source
\beq
{\bar \psi} \gamma_5 \gamma_\mu \psi \to
- \epsilon_{\mu\nu\rho\sigma} \partial_\nu B_{\rho\sigma}
- \frac{i}{4\pi^2} \, \epsilon_{\mu\nu\rho\sigma}
s_\nu \partial_\rho s_\sigma \;.
\label{ultult}
\eeq
In our approach, 
this is a consequence of the anomalous 
behaviour of the fermionic measure under axial gauge transformations and
in this way
the bosonic form of the axial current correctly yields its
anomalous divergence.  
We also mention the possibility of considering the particular case
of a purely chiral external source ($s_\mu \equiv \pm t_\mu$), and obtaining
a bosonized version for this model. The Kalb-Ramond field then
corresponds to a particular `chiral' combination of $A$ and $B$,
namely $A_{\mu\nu} \pm B_{\mu\nu}$. 

As stressed above, recipes (\ref{ult}) and (\ref{ultult}) can be considered
exact apart from the fact that if one is to work
in the bosonic version one has to
use an approximate expression for the bosonic action. The
one we proposed is based in a quadratic approximation and
leads to the bosonic generating functional
presented in eqs.(\ref{boson})-(\ref{chu}).

It should be noted that, besides playing an important role in 
the bosonization rule for the axial current, the chiral
Jacobian also
affects the actual form of the bosonized action,
being the cause of the existence of non-quadratic terms
in the currents, and of the coupling between the field
strength for the Kalb-Ramond field $B$ and the currents. 
This situation may be contrasted with the one of having just
a vector current, where all the non-quadratic terms disappear
if the fermionic determinant is evaluated up to second order
in the fields, as we did.
These non-quadratic terms are a signal of the
anomalous Ward identity linking the divergence of
the axial current and two vector currents. In spite of the fact that we
haven't included the triangle diagram in our approximate evaluation
of the fermionic determinant, partial information from it has
shown up from the Jacobian whose
calculation implies the knowledge of the exact axial anomaly.
It is interesting to note that, as it happens for the complete
$d=2$ bosonization recipe, the axial anomaly determines 
basic properties of $d=4$ bosonization (For odd dimensional
spaces it is the parity anomaly which seems to play a
similar role \cite{FS}-\cite{FAS}).

Finally, a comment about the choice of the actual
value of the renormalized `mass term' for the axial source $s_\mu$:
If the source is not dynamical, there is no propagator associated
to it and, needless to say, its natural renormalized value is 
zero, since this value will not be modified by any other higher
order correction (they would require diagrams with internal $s_\mu$
lines). When the source is dynamical, the actual value of the renormalized
mass is arbitrary and becomes a new quantity to be `measured'.
A model with dynamics for $s_\mu$ is not necessarily anomalous, one
may consider a system of many fermionic species, with their
charges adjusted in order to cancel non-trivially the anomaly. It
should be emphasized that if the vector source is not dynamical, the
regularization can be chosen differently since the 
criterion of preserving the conservation of the vector current at
the quantum level no longer applies.

\medskip
\underline{Acknowledgements}.  This work was supported in part by
Fundaci\'on Antorchas.  F.A.S. is partially
supported by a Commission of the European Communities contract
No:C11*-CT93-0315.
CDF thanks the Universidad Nacional de La Plata
for its kind hospitality.
\newpage

%
\end{document}